\documentclass[conference]{IEEEtran}
\IEEEoverridecommandlockouts
\usepackage{cite}
\usepackage{amsmath,amssymb,amsfonts}
\usepackage{algorithmic}
\usepackage{graphicx}
\usepackage{textcomp}
\usepackage{xcolor}
\begin{document}
	
	\title{VddNet: Vine Disease Detection Network Based on Multispectral Images and Depth Map}

	\author{\IEEEauthorblockN{Mohamed Kerkech}
		\IEEEauthorblockA{\textit{INSA-CVL, Univ. Orl\'{e}ans,} \\
			\textit{PRISME, EA 4229, F18022,}\\
			Bourges, France \\
			mohamed.kerkech@insa-cvl.fr}
		\and
		\IEEEauthorblockN{Adel Hafiane}
		\IEEEauthorblockA{\textit{INSA-CVL, Univ. Orl\'{e}ans,} \\
			\textit{PRISME, EA 4229, F18022,}\\
			Bourges, France \\
			adel.hafiane@insa-cvl.fr}
		\and
		\IEEEauthorblockN{Raphael Canals}
		\IEEEauthorblockA{\textit{Univ. Orl\'{e}ans, INSA-CVL,} \\
			\textit{PRISME, EA 4229, F45072,}\\
			Orl\'{e}ans, France \\
			raphael.canals@univ-orleans.fr}
	}
	
	\maketitle
	
	\begin{abstract}
		Early detection of vine disease is important to avoid spread of virus or fungi. Disease propagation can lead to a huge loss of grape production and disastrous economic consequences, therefore the problem represents a challenge for the precision farming. In this paper, we present a new system for vine disease detection. The article contains two contributions: the first one is an automatic orthophotos registration method from multispectral images acquired with an unmanned aerial vehicle (UAV). The second one is a new deep learning architecture called VddNet (Vine Disease Detection Network). The proposed architecture is assessed by comparing it with the most known architectures: SegNet, U-Net, DeepLabv3+ and PSPNet. The deep learning architectures were trained on multispectral data and depth map information. The results of the proposed architecture show that the VddNet architecture achieves higher scores than the base line methods. Moreover, this study demonstrates that the proposed system has many advantages compared to methods that directly use the UAV images.
	\end{abstract}
	
	\begin{IEEEkeywords}
		Plant disease detection, precision agriculture, UAV multispectral images, machine learning, orthophotos registration, 3D information, orthophotos segmentation.
	\end{IEEEkeywords}
	
	\section{Introduction}
	In agricultural fields, the main causes of losing quality and yield of harvest are virus, bacteria, fungi and pest~\cite{Oerke2006}. Against these harmful pathogens, farmers generally treat the global crop to prevent different diseases. However, using large amount of chemicals has a negative impact on human health and ecosystems. This constitutes a significant problem to be solved; the precision agriculture presents an interesting alternative.
	
	In recent decades, the precision agriculture~\cite{Patricio2018,Mogili2018} has introduced many new farming methods to improve and optimize crop yields: it constitutes a research field in continuous evolution. New  sensing technologies and algorithms have enabled the development of several applications such as water stress detection~\cite{Bellvert2014}, vigour evaluation~\cite{Mathews2014}, estimation of evaporate-transpiration and harvest coefficient~\cite{Vanino2015}, weeds localization~\cite{Bah2020,DianBah2018}, disease detection~\cite{Tichkule2016,Pinto2017}, etc. 
	
	Disease detection in vine is an important topic in precision agriculture~\cite{macdonald_remote_2016-1, Junges2018,gennaro_unmanned_2016,Albetis2017, Albetis2019,Al-Saddik2017,Al-saddik2018,Al-Saddik2019,Rancon2019, Kerkech2018,Kerkech2019,Kerkech2020}. The aim is to detect and treat the infected area at the right place, and the right time and with the right dose of phytosanitary products. At early stage, it is easier to control diseases with small amounts of chemical products. Indeed, intervention before infection spreads offers many advantages such as: preservation of vine, grap production and environment, and reducing the economics losses. To achieve this goal, frequent monitor of the parcel is necessary. Remote sensing (RS) methods are among the most widely used for that purpose and became essential in the precision agriculture. RS images can be obtained at leaf or parcel scale. At the leaf level, images are acquired using a photo sensor either held by a person~\cite{Singh2017} or mounted on a mobile robot~\cite{Pilli2015}. At the parcel level, satellite was the standard RS imaging system~\cite{Abbas2020,Ulabhaje2018}. Recently, drones or UAVs have gained popularity due to their low cost, high resolution images, flexibility, customization, easy data access~\cite{Mukherjee2019}. In addition, unlike satellite imaging, UAV does not have the cloud problem, which has helped to solve many remote sensing problems.
	
	Parcels monitoring generally requires orthophotos building from geo-referenced visible and infrared UAV images. However, two separated sensors generate a spatial shift between images of  the two sensors. This problem also occurred after building the orthophotos. It has been established that it is more interesting to combine the information from the two sensors to increase the efficiency of disease detection. Therefore, images registration is required.
	
	The existent algorithms of registration rely on an approach based on either the area or feature methods. The most commonly used ones in the precision agriculture are feature-based methods, which are based on matching features between images~\cite{Xiong2010}. In this study, we adopted the feature-based approach to align orthophotos of the visible and infrared ranges. Then, the two are combined for the disease detection procedure, where the problem consists in assigning a class-label to each pixel. For that purpose, the deep learning approach is nowadays the most preferred approach for solving this type of problem.
	
	Deep learning methods~\cite{Unal2020} have achieved a high level of performance in many applications in which different network architectures have been proposed. For instance, R-CNN~\cite{Girshick2014}, Siamese~\cite{Bertinetto2016}, ResNet~\cite{He2016}, SegNet~\cite{Badrinarayanan} are architectures respectively used for object detection, tracking, classification, segmentation which operate in most cases in visible ranges. However, in certain situations, the input data are not only visible images but can be combined with multispectral or hyperspectral images~\cite{Polder2019}, and even depth information~\cite{Naseer2019}. In these contexts, the architectures can undergo modification for improving the methods~\cite{Sa2017}. Thus, in some studies~\cite{Ren2017,Gene-Mola2019,Bezen2020,Aghi2020}, depth information is used as input data. These data generally provide precious information about scene or environment.
	
	Depth or height information is extracted from the 3D reconstruction or photogrammetry processing. In UAV remote sensing imagery, the photogrammetry processing allows to build a digital surface model (DSM) before creating the orthophoto. The DSM model can provide many information about the parcel, such as the land variation and objects on its surface. Certain research works have showed the ability to extract vinerows by generating a depth map from the DSM model~\cite{Burgos2015,Matese2017,Weiss2017}. These solutions have been proposed to solve the vinerows misextraction resulting from the NDVI vegetation index. Indeed, in some situations, the NDVI method cannot be used to extract vinerows when the parcel has a green grassy soil. The advantage of the depth map is the ability to separate areas above-ground from the ground, even if the color is the same for all zones. So far, there has been no work on the vine disease detection that combines depth and multispectral information with a deep learning approach.
	
	This paper presents a new system for vine disease detection using multispectral UAV images. It combines a high accurate orthophotos registration method, a depth map extraction method and a deep learning network adapted to the vine disease detection data. 
	
	The article is organized as follows. Section \ref{RelatedWork} presents a review of related works. Section \ref{MaterialMethod} describes the materials and methods used in this study. Section \ref{Experimentation} details the experiments. Section \ref{Discussion} discusses performances and limitations of the proposed method. Finally, section \ref{Conclusion} concludes the paper and introduces ideas to improve the method.
	
	\begin{figure*}[!h]
	\centering{\includegraphics[width=\textwidth]{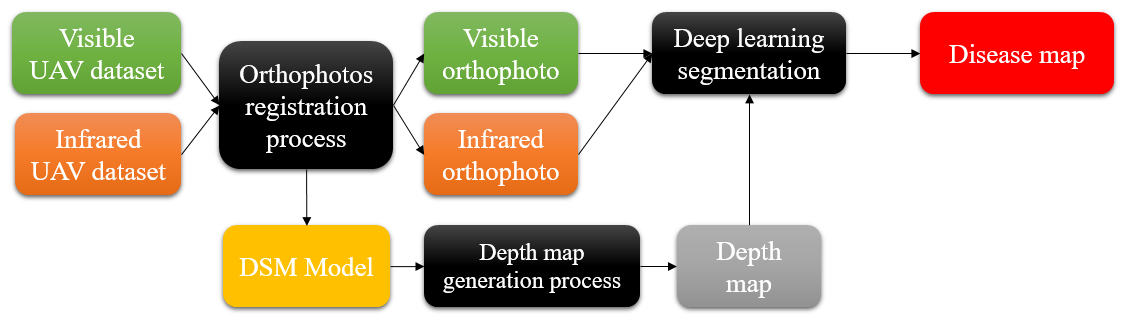}}
	\caption{\label{Overview} The proposed vine disease detection system.}	
    \end{figure*}	
	
	\section{Related work}
	\label{RelatedWork}
	Plant disease detection is an important issue in precision agriculture. Many researches have been carried out and a large survey has been realised by Mahlein (2016)~\cite{mahlein_plant_2016}, Kaur et al. (2018)~\cite{Kaur2019} , Saleem et al. (2019)~\cite{Saleem2019}, Sandhu et al. (2019)~\cite{Sandhu2019} and Loey et al. (2020)~\cite{Loey2020}. Schor et al. (2016)~\cite{Schor2016} presented a robotic system for detecting powdery mildew and wilt virus in tomato crops. The system is based on RGB sensor mounted on a robotic arm. Image processing and analysis were developed using the principal component analysis and the coefficient of variation algorithms. Sharif et al. (2018)~\cite{Sharif2018} developed a hybrid method for disease detection and identification in citrus plants. It consists in the lesion detection on the citrus fruits and leaves, followed by a classification of the citrus diseases. Ferentinos (2018)~\cite{Ferentinos2018} and Argüeso et al. (2020)~\cite{Argueso2020} built a CNN model to perform plant diagnosis and disease detection using images of plant leaves. Jothiaruna et al. (2019)~\cite{Jothiaruna2019} proposed a segmentation method for the disease detection at the leaf scale using a color features and region growing method. Pantazi et al. (2019)~\cite{Pantazi2019} presented an automated approach for crop disease identification on images of various leaves. The approach consists in using a local binary patterns algorithm for extracting features and performing classification into disease classes. Abdulridha et al. (2019)~\cite{Abdulridha2019a} proposed a remote sensing technique for the early detection of avocado diseases. Hu et al. (2020)~\cite{Hu2020} combined an internet of things (IoT) system with deep learning to create a solution for automatically detecting various crop diseases and communicating the diagnostic results to farmers.
	
	Disease detection in vineyards has been increasingly studied in recent years~\cite{macdonald_remote_2016-1, Junges2018,gennaro_unmanned_2016,Albetis2017, Albetis2019, Al-Saddik2017, Al-saddik2018, Al-Saddik2019, Rancon2019, Kerkech2018,Kerkech2019,Kerkech2020}. Some works are realised at the leaf scale, and others at the crop scale. MacDonald et al. (2016) ~\cite{macdonald_remote_2016-1} used a Geographic Information System (GIS) software and multispectral images for detecting the leafroll-associated virus in vine. Junges et al. (2018)~\cite{Junges2018} investigated vine leaves affected by the esca in hyperspectral ranges and di Gennaro et al. (2016)~\cite{gennaro_unmanned_2016} worked at the crop level (UAV images). Both studies concluded that the reflectance of healthy and diseased leaves are different. Albetis et al. (2017)~\cite{Albetis2017} studied the Flavescence dorée detection in UAV images. The results obtained showed that the vine disease detection using aerial images is feasible. The second study of Albetis et al. (2019)~\cite{Albetis2019} examined of the UAV multispectral imagery potential in the detection of symptomatic and asymptomatic vines. Al-Saddik has conducted three studies on vine disease detection using hyperspectral images at the leaf scale. The aim of the first one (Al-Saddik et al. 2017)~\cite{Al-Saddik2017} was to develop spectral disease indices able to detect and identify the Flavescence dorée on grape leaves. The second one (Al-Saddik et al. 2018)~\cite{Al-saddik2018} was performed to differentiate yellowing leaves from leaves diseased by esca through classification. The third one (Al-saddik et al., 2019)~\cite{Al-Saddik2019} consisted in determining the best wavelengths for the detection of the Flavescence dorée disease. Rançon et al. (2019)~\cite{Rancon2019} conducted a similar study for detecting esca disease. Image sensors were embedded on a mobile robot. The robot moved along the vinerows to acquire images. To detect esca disease, two methods were used: the scale Invariant Feature Transform (SIFT) algorithm and the MobileNet architecture. The authors concluded that the MobileNet architecture provided a better score than the SIFT algorithm. In the framework of previous works, we have realized three studies on vine disease detection using UAV images. The first one (Kerkech et al. 2018)~\cite{Kerkech2018} was devoted to esca disease detection in the visible range using the LeNet5 architecture combined with some color spaces and vegetation indices. In the second study (Kerkech et al. 2019)~\cite{Kerkech2019}, we used near-infrared images and visible images. Disease detection was considered as a semantic segmentation problem performed by the SegNet architecture. Two parallel SegNet were applied for each imaging modality and the results obtained were merged to generate a disease map. In (Kerkech et al. 2020)~\cite{Kerkech2020}, a correction process using a depth map was added to the output of the previous method. A post-processing with these depth information demonstrated the advantage of this approach to reduce detection errors.

	\section{Materials and methods}
	\label{MaterialMethod}
	This section presents, the materials and each component of the vine disease detection system. Figure~\ref{Overview} provides an overview of the methods. It includes the following steps: data acquisition, orthophotos registration, depth map building and orthophotos segmentation (disease map generation). The next sections detail these different steps.
	
	\begin{figure*}[!h]
	\centering{\includegraphics[width=\textwidth]{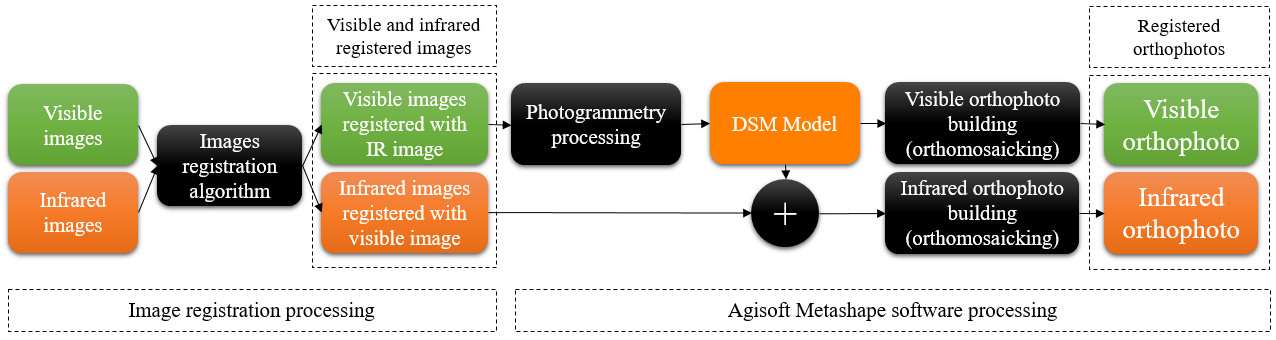}}
	\caption{\label{Orthophoto_Registration} The proposed orthophotos registration method.}	
	\end{figure*}	
	
	\subsection{Data acquisition}
	Multispectral images are acquired using a quadricopter UAV that embeds a MAPIR Survey2 camera and a Global Navigation Satellite System (GNSS) module. This camera integrates two sensors in the visible and infrared ranges with a resolution of 16 megapixels (4608$\times$3456 pixels). The visible sensor captures the red, green, and blue (RGB) channels and the infrared sensor captures the red, green, and near-infrared (R-G-NIR) channels. The wavelength of the near-infrared channel is 850 nm. The accuracy of the GNSS module is approximately 1 meter.
	
	The acquisition protocol consists of a drone flying over vines at an altitude of 25 meters and at an average speed of 10 km/h. During flights, the sensors acquire an image every 2 seconds. Each image has a 70\% overlap with the previous and the next ones. Each point of the vineyard has 6 different viewpoints (can be observed on 6 different images). Images are recorded with their GNSS position. Flights are performed at the zenith to avoid shadows, and with moderate weather conditions (light wind and no rain) to avoid UAV flight problems.
	
	\subsection{Orthophotos registration}
	The multispectral acquisition protocol using two sensors causes a shift between visible and infrared images. Hence, a shift in multispectral images automatically implies a shift in orthophotos. Usually, the orthophotos registration is performed manually using the QGIS software. The manual method is time consuming, requires a high focusing to select many key points between visible and infrared orthophotos, and the result is not very accurate. To overcome this problem, a new method for automatic and accurate orthophotos registration is proposed.
	
	The proposed orthophotos registration method is illustrated in Figure~\ref{Orthophoto_Registration} and is divided in two steps. The first one concerns the UAV multispectral images registration and the second permits the building of registered multispectral orthophotos. In this study, the first step uses the optimized multispectral images registration method proposed in~\cite{Kerkech2019}. Based on the AKAZE (Accelerated-KAZE) algorithm, the registration method uses a features matching between visible and infrared images to match key points extracted from the two images and compute the homographic matrix for geometric correction. In order to increase accuracy, the method uses an iterative process to reduce the RMSE (Root Mean Squared Error) of the registration. The second step consists in using the Agisoft Metashape software to build and obtain the registered visible and infrared orthophotos. The Metashape software is based on the Structure from motion (SfM) algorithm for the photogrammetry processing. Building orthophotos requires the UAV images and the digital surface model (DSM). To obtain this DSM model, the software must go through a photogrammetry processing and perform the following steps: alignment of the images to build a sparse point cloud, then a dense point cloud and finally the DSM. The orthophotos building is carried out by the option "build orthomosaic" process in the software. To build the visible orthophotos, it is necessary to use the visible UAV images and the DSM model, while to build a registered infrared orthophoto, it is necessary to use the registered infrared UAV images and the same DSM model of the visible orthophoto. The parameters used in the Metashape software are detailed in Table~\ref{Tab_Photogrammetry_Param}.

	\subsection{Depth map}
	The DSM model previously built in the orthophotos registration process is used here to obtain the depth map. In fact, the DSM model represents the terrain surface variation and includes all objects found here (in this case, objects are vine trees). Therefore, some processings are required to determine only the vine height. To extract the depth map from the DSM, the method proposed in~\cite{Burgos2015} is used. It consists in applying the following processings: the DSM is first filtered using a low-pass filter of size 20 $\times$ 20; this filter is chosen for smoothing the image and to keep only the terrain surface variations also called digital terrain model (DTM). The DTM is thereafter subtracted from the DSM to eliminate the terrain variations and retain only the vine height. Due to the weak contrast of the result, an enhancement processing was necessary. The contrast is enhanced here by using a histogram-based (histogram normalization) method. The obtained result is an image with a good difference in grey levels between vines and non-vines. Once the contrast is corrected, an automatic thresholding using the Otsu’s algorithm is applied to obtain a binary image representing the depth map. 
	
	
	
	\subsection{Segmentation and classification}
	The last stage of the vine disease detection system concerns the data classification. This step is performed using a deep learning architecture for segmentation. Deep learning has proven its performances in numerous research studies and in various domains. Many architectures have been developed, such as SegNet~\cite{Badrinarayanan}, U-Net~\cite{Ronneberger}, DeepLabv3+~\cite{Chen2018a}, PSPNet~\cite{Zhao2017}, etc. Each architecture can provide good results in a specific domain and be less efficient in others. These architectures are generally used for segmentation of complex indoor / outdoor scenes, medical ultrasound images, or even in agriculture. One channel is generally used for greyscale medical imaging or three channels for visible RGB color images. Hence, they are not always adapted to a specific problem. Indeed, for this study, multispectral and depth map data offer additional information. This can improve the segmentation representation and the final disease map result. For this purpose, we have designed our deep learning architecture adapted to the vine disease detection problem, and we have compared it to the most well known deep learning architectures.
	In the following sections, we describe the proposed deep learning architecture and the training process.
	
	\subsubsection{VddNet architecture}
\begin{figure*}[!h]
		\centering{\includegraphics[width=\textwidth]{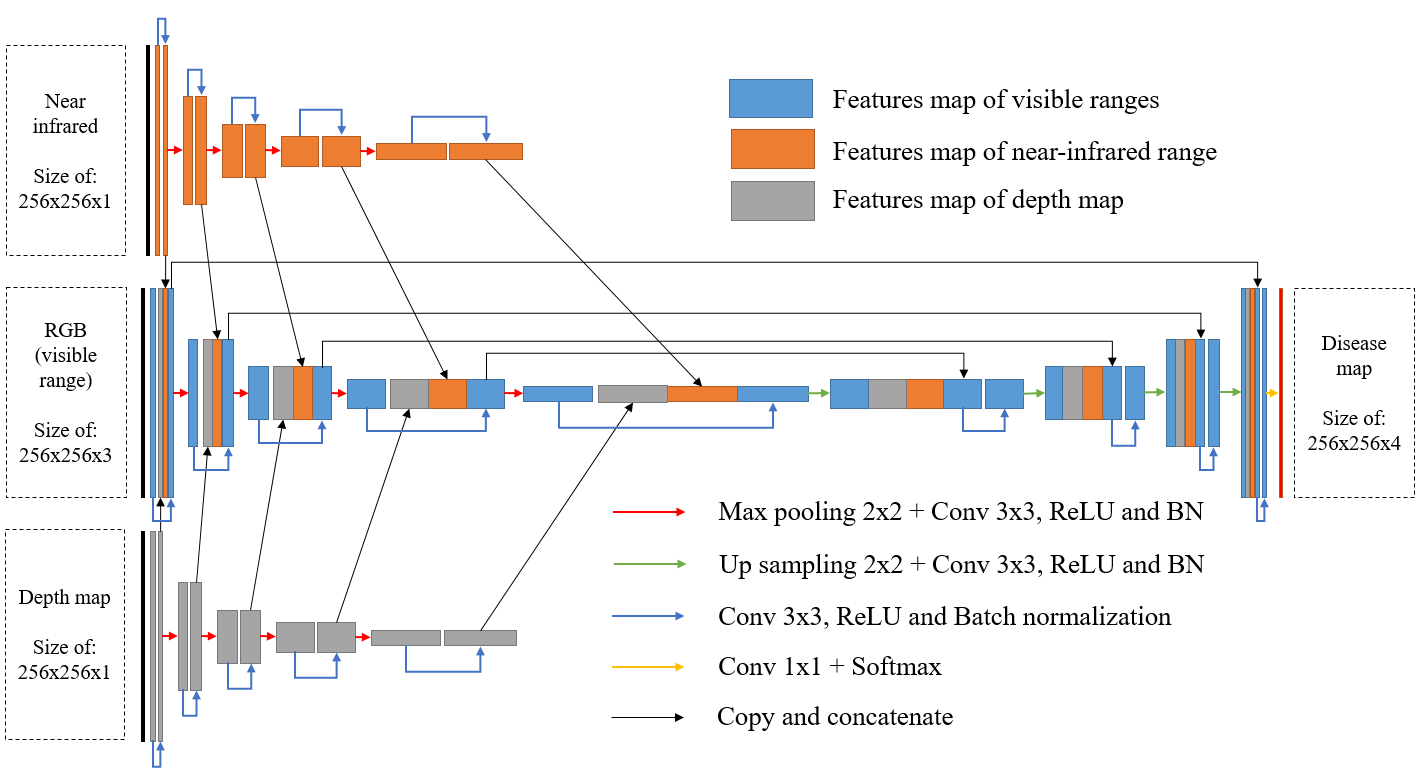}}
		\caption{\label{VddNet_Architecture} VddNet architecture.}	
	\end{figure*}

	Vine Disease Detection Network (VddNet), Figure~\ref{VddNet_Architecture} is inspired by VGG-Net~\cite{Sermanet2014}, SegNet~\cite{Badrinarayanan}, U-Net~\cite{Ronneberger} and the parallel architectures proposed in~\cite{Ren2017,Liu2017,Adhikari2019,Dunnhofer2020}. VddNet is a parallel architecture based on the VGG encoder, it has as inputs three types of data: visible a RGB image, a near-infrared image and a depth map. VddNet is dedicated to segmentation, so, the output has the same input, with a number of channels equal to the number of classes (4). It is designed with three parallel encoders and one decoder. Each encoder can typically be considered as a convolutional neural network without the fully connected layers. The convolutional operation is repeated twice using a 3$\times$3 mask, a rectified linear unit (ReLU), a batch normalization and a subsampling using a max pooling function of 2$\times$2 size and a stride of 2. The number of features map channels is doubled at each subsampling step. The idea of VddNet is to encode each type of data separately and at the same time concatenate the near-infrared and the features map of the depth map with the visible features map before each subsampling. Hence, the central encoder preserves the features of the near-infrared and the depth map data merged with the visible features map, and concatenated at the same time. The decoder phase consists of upsampling and convolution with a 2$\times$2 mask. It is then followed by two convolution layers with a 3$\times$3 mask, a rectified linear unit, and a batch normalization. In contrast to the encoder phase, after each upsampling operation, the number of features map channels is halved. Using the features map concatenation technique of near-infrared and depth map, the decoder retrieves features lost during the merging and the subsample process. The decoder follows the same steps until it reaches the final layer, which is a convolution with a 1$\times$1 mask and a softmax providing classes probabilities, at pixel-wise.
	
	\subsubsection{Training dataset}
	To build the training dataset, four steps are required: data source selection, classes definition, data labelling, and data augmentation.
	
	The first step is probably the most important one. Indeed, to allow a good learning, the data source for feeding models must represent the global data in terms of richness, diversity and classes. In this study, a particular area was chosen that contains a slight shadow area, brown ground (soil) and a vine partially affected by mildew.
	
	Once the data source has been selected, it is necessary to define the different classes present in these data. For that purpose, each type of data (visible, near-infrared and depth map) is important in this step. In visible and near-infrared images, four classes can be distinguished. On the other hand, the depth map contains only two distinct classes which are the vine canopy and the non-vine. Therefore, the choice of classes must match all data types. Shadow is the first class; it is any dark zone. It can be either on the vine or on the ground. This class was created to avoid confusion and misclassification on a non-visible pattern. Ground is the second class; From one parcel to another, ground is generally different. Indeed, the ground can have many colors such as brown, green, grey, etc. To solve this color confusion, the ground is chosen as any pixels in the non-vine zone from the depth map data. Healthy vine is the third class; it is the green leaves of the vine. Usually it is easy to classify this data, but when ground is also green, this leads to confusion between vine and ground in 2D images. To avoid that, the healthy class is defined as the green color in the visible spectrum and belonging to the vine canopy according to the depth map. The fourth and last class corresponds to diseased vine. Disease symptoms can present several colors in the visible range such as yellow, brown, red, golden, etc. In the near-infrared, it is only possible to differentiate between healthy and diseased reflectances. In general, diseased leaves have a different reflectance than healthy leaves~\cite{Al-saddik2018}, but some confusion between disease and ground classes may occur when the two colors are similar. Ground must also be eliminated from the disease class using the depth map.
	
	Data labelling was performed with the semi-automatic labelling method proposed in~\cite{Kerkech2019}. The method consists in using automatic labelling in a first step, followed by manual labelling in a second step. The first step is based on the deep learning LeNet-5~\cite{LeCun1998} architecture, where the classification is carried out using a 32$\times$32 sliding window and a 2$\times$2 stride. The result is equivalent to a coarse image segmentation which contains some misclassifications. To refine the segmentation, output results were manually corrected using the Paint.Net software. This task was conducted based on the ground truth (realized in the crop by a professional reporting occurred diseases), and observations in the orthophotos.
	
	The last stage is the generation of a training dataset from the labelled data. In order to enrich the training dataset and avoid an overfitting of networks, data augmentation methods~\cite{Dellana2016} are used in this study. A 256$\times$256 pixels patches dataset is generated from the data source matrix and its corresponding labelled matrix. The data source consists of multimodal and depth map data and has a size of 4626$\times$3904$\times$5. Four data augmentation methods are used: translation, rotation, under and oversampling, and brightness variation. Translation was performed with an overlap of 50\% using a sliding window in the horizontal and vertical displacements. The rotation angle was set at $ 30^\circ $, $ 60^\circ $ and $ 90^\circ $. Under and oversampling were parametrized to obtain 80\% and 120\% of the original data size. Brightness variation is only applied to multispectral data. Pixel values are multiplied by the coefficients of 0.95 and 1.05 which introduce a brightness variation of $ \pm $ 5\%. Each method brings an effect on the data (translation, rotation ...) allowing the networks to learn respectively transition, vinerows orientations, acquisition scale variation and weather conditions. At the end, the data augmentation generated 35.820 patches.

	\section{Experimentations and results}
	\label{Experimentation}
	This section presents the different experimental devices, as well as qualitative and quantitative results. The experiments are performed on Python 2.7 software, using the Keras 2.2.0 library for the development of deep learning architectures, and GDAL 3.0.3 for the orthophotos management. The Agisoft Metashape software version 1.6.2 is also used for photogrammetry processing. The codes were developped under the Linux Ubuntu 16.04 LTS 64-bits operating system and run on a hardware with an Intel Xeon 3.60 GHz $\times$ 8 processor, 32 GB RAM, and a NVidia GTX 1080 Ti graphics card with 11 GB of internal RAM. The cuDNN 7.0 library and the CUDA 9.0 Toolkit are used for deep learning processing on GPU.

	\begin{table*}[!h]
		\normalsize
		\renewcommand{\arraystretch}{1.25}
		\caption{\label{Tab_Photogrammetry_Param} The parameters used for the orthophotos building process in the Agisoft Metashape software.}	
		\begin{center}	
			\begin{tabular}{|l | c |} 
				\hline
				\underline{Sparse point cloud} & \\
				{	Accuracy :} & High\\
				{	Image pair selection :} & Ground control\\
				{	Constrain features by mask :} & No\\
				{	Maximum number of feature points :} & 40,000\\
				&\\
				\underline{Dense point cloud} &\\
				{	Quality :} & High\\
				{	Depth filtering :} & Disabled\\
				&\\			
				\underline{Digital Surface Model} &\\
				{	Type :} & Geographic\\
				{	Coordinate system :} & WGS 84 (EPSG::4326)\\
				{	Source data :} & Dense cloud\\
				&\\
				\underline{Orthomosaic} &\\
				{	Surface :} & DSM\\
				{	Blending mode :} & Mosaic\\
				\hline
			\end{tabular}
		\end{center}
	\end{table*}

	\subsection{Orthophotos registration and depth map building}
	To realize this study, multispectral and depth map orthophotos were required. Two parcels were selected and data were aquired at two different times to construct the orthophotos dataset. Each parcel had one or more of the following characteristics: with or without shadow, green or brown ground, healthy or partially diseased. Registered visible and infrared orthophotos were built from multispectral images using the optimized image registration algorithm~\cite{Kerkech2019} and the Agisoft Metashape software version 1.6.2. Orthophotos were saved in the geo-referenced file format "TIFF". The parameters used in the Metashape software are listed in Table~\ref{Tab_Photogrammetry_Param}. 

	To evaluate the registration and depth maps quality, we chosed chessboard test pattern. Figure~\ref{QR_Orthophoto_Registration} presents an example of visible and infrared orthophotos registration. As it can be seen, the alignment between the two orthophotos is accurate. The registration of the depth map with the visible range also provides good results (Figure~\ref{QR_DM_Chessboard}).
	
	\subsection{Training and testing architectures}
	In order to determine the best parameters for each deep learning architecture, four cross-optimizers with two loss functions were compared. Architectures were compiled using either the loss function "cross entropy" or "mean squared error", and with one among the four optimizers: SGD~\cite{Hoffman2013}, Adadelta~\cite{Zeiler2012}, Adam~\cite{Kingma2015}, or Adamax~\cite{Zeng2016}. Once the best parameters were defined for each architecture, a final fine tuning was performed on the "learning rate" parameter to obtain the best results (to achieve a good model without overfitting). The best parameters found for each architecture are presented in Table~\ref{Tab_DL_Param}.
	
	\begin{table*}[!h]
		\normalsize
		\renewcommand{\arraystretch}{1.25}
		\caption{\label{Tab_DL_Param} The parameters used for the different deep learning architectures.}	
		\begin{center}	
			\begin{tabular}{|c | c | c | c | c|} 
				\hline
				Network&Base model&Optimizer&Loss function&Learning rate\\
				\hline
				SegNet&VGG-16&Adadelta&{Categorical cross entropy}&1.0\\
				\hline
				U-Net&VGG-11&SGD&Categorical cross entropy&0.1\\
				\hline
				PSP-Net&ResNet-50&Adam&Categorical cross entropy&0.001\\
				\hline
				DeepLabv3+&Xception&Adam&Categorical cross entropy&0.001\\
				\hline
				VddNet&Parallel VGG-13&SGD&Categorical cross entropy&0.1\\
				\hline														
			\end{tabular}
		\end{center}
	\end{table*}
	
	For training the VddNet model, data from visible, near-infrared and depth maps were incorporated separately in the network inputs. For the others architectures, a multi-data matrix consists of 5 channels with a size of 256$\times$256. The first 3 channels correspond to the visible spectrum, the 4th channel for the near-infrared data and the 5th channel for the depth map. Each multi-data matrix has a corresponding labelled matrix. Models training is an iterative process that is fixed at 30.000 epochs for each model. For each iteration, a batch of 5 multi-data matrices with their corresponding labelled matrices are randomly selected from the dataset and sent to feed the model. In order to check the convergence of the model, a test using validation data is performed each 10 iterations.
	
\begin{figure*}[!h]
	\centering{\includegraphics[width=\textwidth]{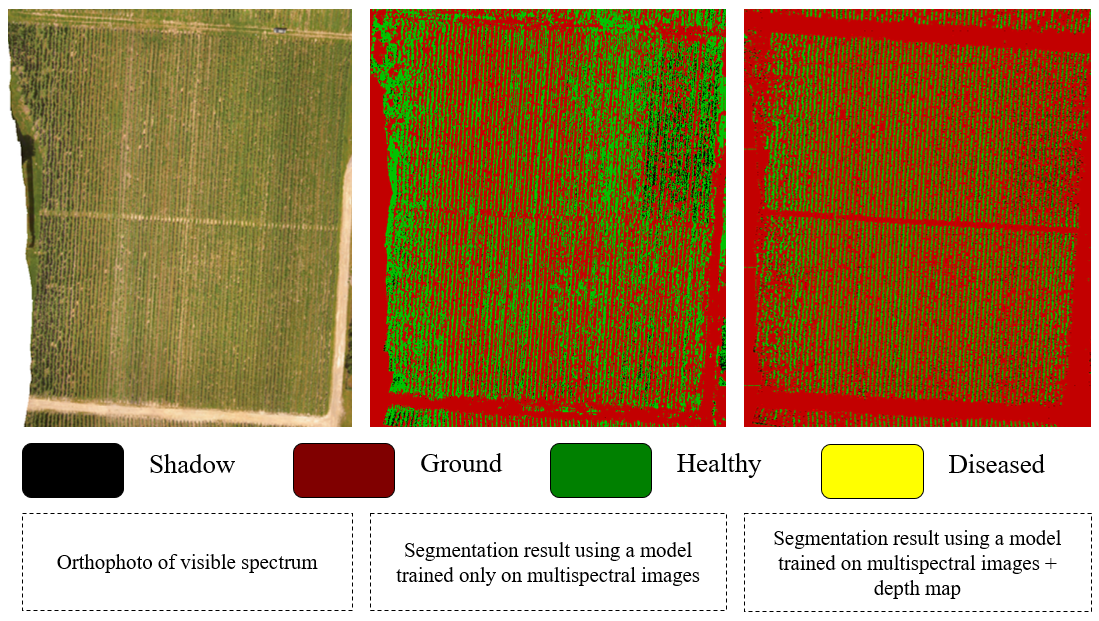}}
	\caption{\label{Difference_Between_Training_1} Difference between a SegNet model trained only on multispectral data and the same trained on multispectral data combined with depth map information. The presented example is on a orthophoto of healthy parcel with a green ground.}	
\end{figure*}

\begin{figure*}[!h]
	\centering{\includegraphics[width=\textwidth]{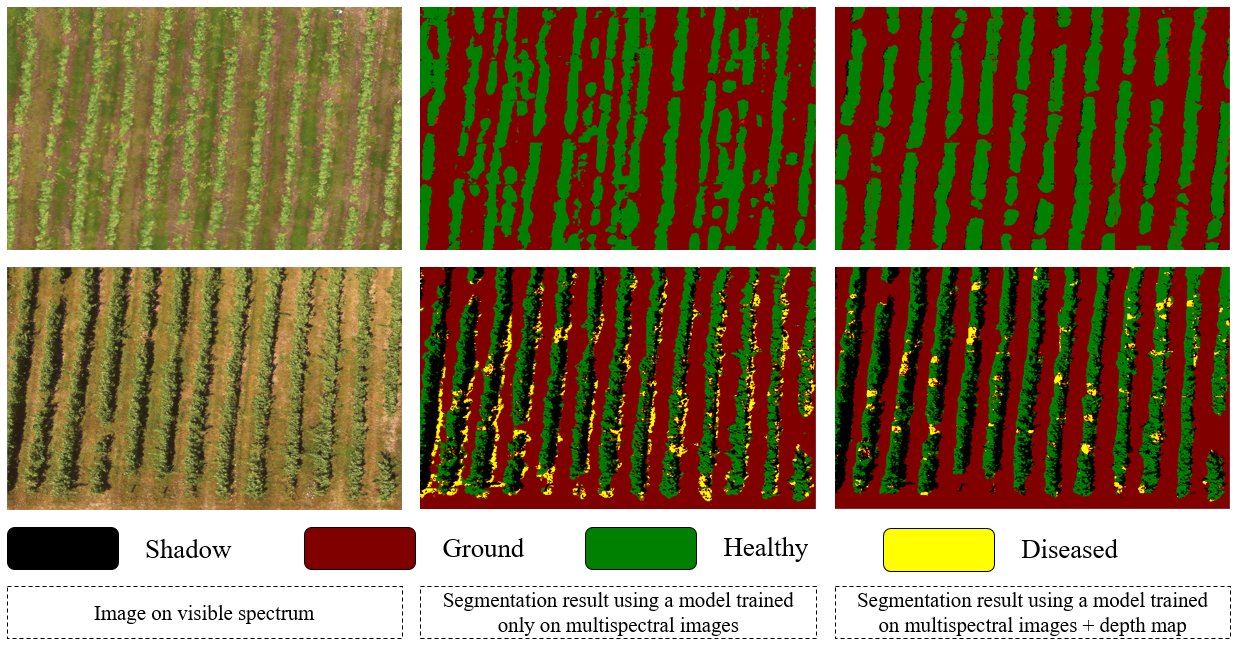}}
	\caption{\label{Difference_Between_Training_2} Difference between a SegNet model trained only on multispectral data and the same trained on multispectral data combined with depth map information. Two examples are presented here, the first row is an example on a healthy parcel with a green ground. The second one is an example on a partially diseased parcel with a brown ground.}	
\end{figure*}

	A qualitative study was conducted for determining the importance of depth map information. For this purpose, an experience was conducted by training the deep learning models with only multispectral data and with a combination of both (multispectral and depth maps). The comparison results are shown in Figures~\ref{Difference_Between_Training_1} and~\ref{Difference_Between_Training_2}.
	
	To test the deep learning models, test areas are segmented using a 256$\times$256 sliding window (without overlap). For each position of the sliding window, the visible, near-infrared and depth maps are sent to the networks inputs (respecting the data order for each architecture) in order to perform segmentation. The output of the networks is a matrix of size of 256$\times$256$\times$4. Results are saved after an application of the Argmax function. They are then stitched together to obtain the original size of the orthophoto tested data.

	\subsection{Segmentation performance measurements}
	Segmentation performance measurements are expressed in terms of using recall, precision, F1-Score/Dice and accuracy (using equations~\ref{Recall},~\ref{Precision},~\ref{F1Score},~\ref{Dice} and~\ref{ACC}) for each class (shadow, ground, healthy and diseased) at grapevine-scale. Grapevine-scale assessment was chosen because pixel-wise evaluation is not suitable to provide disease information. Moreover, imprecision of the ground truth, small surface of the disease and difference of deep learning segmentation results do not allow a good evaluation of the different architectures, at pixel-wise. These measurements use a sliding window equivalent to the average size of a grapevine (in this study, approximatively 64$\times$64 pixels). For each step of the sliding window, the class evaluated is the dominant class in the ground truth. The window is considered "true positive" if the dominant class is the same as the ground truth, otherwise it is a "false positive". The confusion matrix is updated for each step. Finally, the score is given by: 
	
	\begin{equation}
	\label{Recall}
	\begin{array}{c}
	$$Recall = \frac{TP}{TP+FN}$$
	\end{array}
	\end{equation}
	
	\begin{equation}
	\label{Precision}
	\begin{array}{c}
	$$Precision = \frac{TP}{TP+FP}$$
	\end{array}
	\end{equation}
	
	\begin{equation}
	\label{F1Score}
	\begin{array}{c}
	$$F1-Score = 2 \frac{Recall \times Precision}{Recall+Precision} = \frac{2TP}{FP+2TP+FN}$$
	\end{array}
	\end{equation}
	
	\begin{equation}
	\label{Dice}
	\begin{array}{c}
	$$Dice = \frac{2 |X \cap Y|}{|X|+|Y|} = \frac{2(TP)}{(FP+TP)+(TP+FN)} = \frac{2TP}{FP+2TP+FN}$$
	\end{array}
	\end{equation}
	
	\begin{equation}
	\label{ACC}
	\begin{array}{c}
	$$Accuracy = \frac{TP+TN}{TP+TN+FP+FN}$$
	\end{array}
	\end{equation}
	
	where TP, TN, FP and FN are the number of samples for "true positive", "true negative", "false positive" and "false negative" respectively. Dice equation is defined by X (set of ground truth pixels) and Y (set of the classified pixels).

	\begin{figure*}[!h]
	\centering{\includegraphics[width=\textwidth]{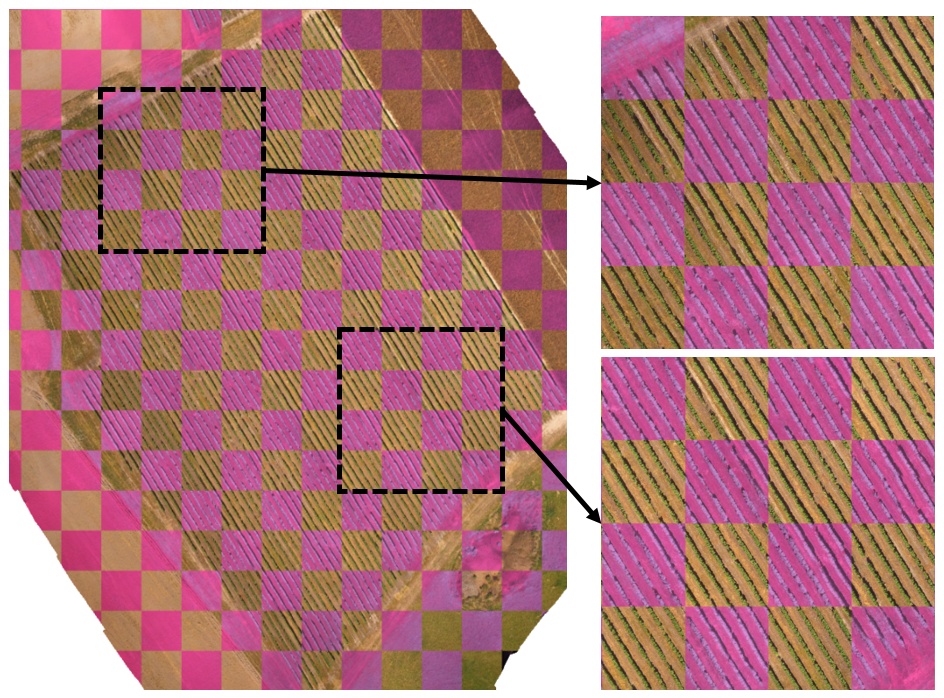}}
	\caption{\label{QR_Orthophoto_Registration} Qualitative results of orthophotos registration using a chessboard pattern.}	
\end{figure*}

\begin{figure*}[!h]
	\centering{\includegraphics[width=\textwidth]{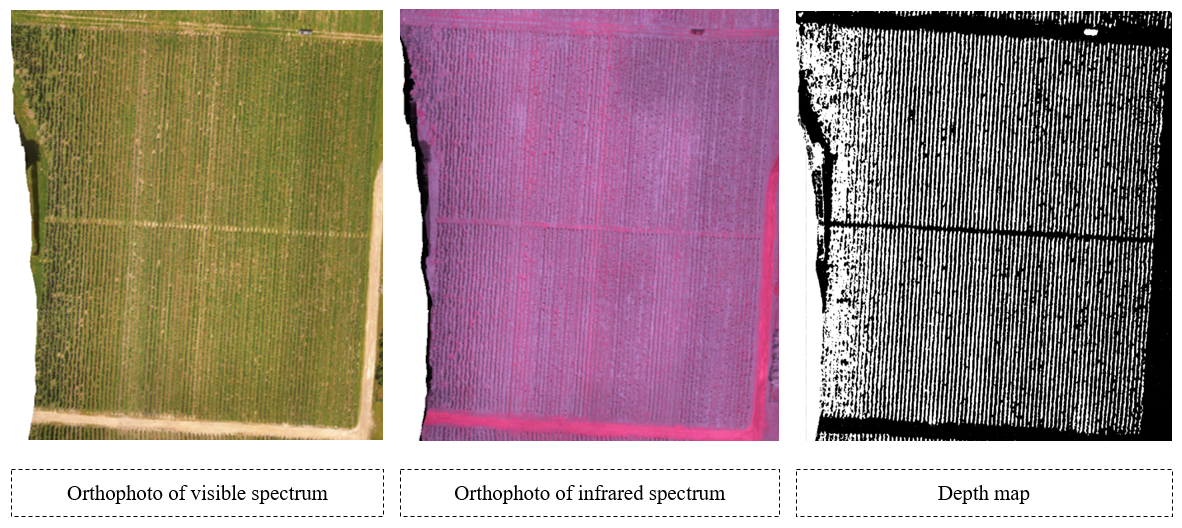}}
	\caption{\label{QR_Orthophotos_DM} Qualitative results of orthophotos and depth map.}	
\end{figure*}

\begin{figure*}[!h]
	\centering{\includegraphics[width=\textwidth]{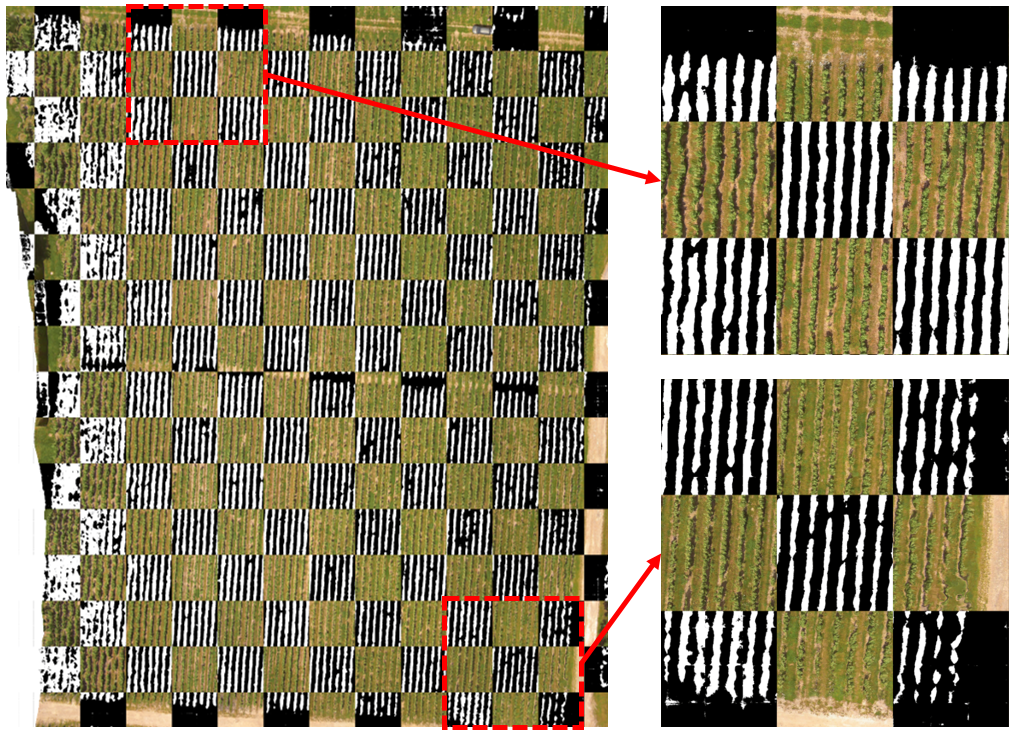}}
	\caption{\label{QR_DM_Chessboard} Evaluation of the depth map alignment using a chessboard pattern.}	
\end{figure*}

	\begin{figure*}[!h]
	\centering{\includegraphics[width=\textwidth]{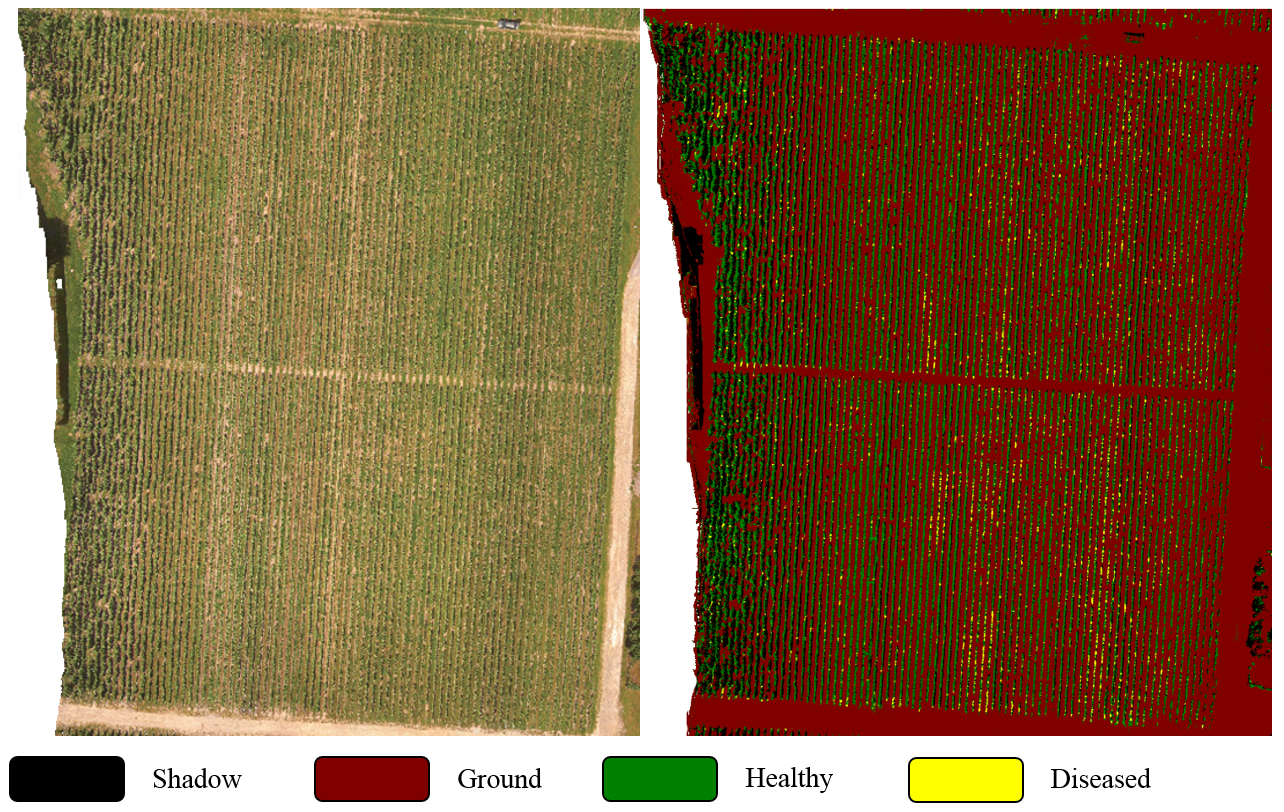}}
	\caption{\label{QR_Orthophoto_VddNet_1} Qualitative result of VddNet on a parcel partially contaminated with mildew and with green ground. The visible orthophoto of the healthy parcel is in the left side, and its disease map in the right side.}	
\end{figure*}

\begin{figure*}[!h]
	\centering{\includegraphics[width=\textwidth]{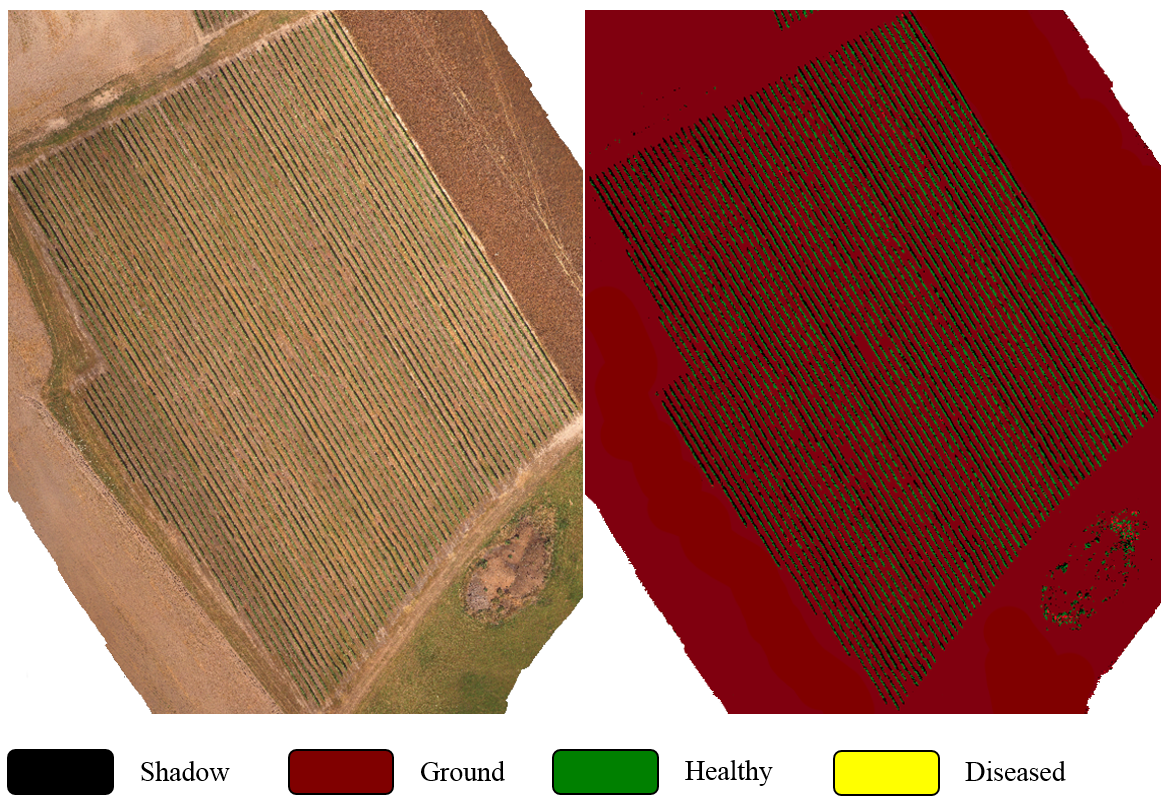}}
	\caption{\label{QR_Orthophoto_VddNet_2} Qualitative result of VddNet on a healthy parcel with brown ground. The visible orthophoto of the healthy parcel is in the left side, and its disease map in the right side.}	
\end{figure*}

	\section{Discussion}
	\label{Discussion}
	To validate the proposed vine disease detection system, it is necessary to evaluate and compare qualitative and quantitative results for each block of the whole system. For this purpose, several experiments were conducted at each step of the disease detection procedure. The first experience was carried out on the multimodal orthophotos registration. Figure~\ref{QR_Orthophoto_Registration} shows the obtained results. As can be seen, the continuity of the vinerows is highly accurate and the continuity is respected between the visible and infrared ranges. However, if image acquisition  is incorrectly conducted, this results in many registration errors. To avoid these problems, two rules must be followed. The first one is the overlapping between visible and infrared images acquired in the same position, which must be greater than 85\%. The second rule is that the overlapping between each acquired image must be greater than 70\%; this rule must be respected in both ranges. Non-compliance with the first rule affects the building of the registered infrared orthophoto. Indeed, this latter may present some black holes (this means that there is no data available to complete theses holes). Non-compliance with the second rule affects the photogrammetry processing and the DSM model. This can lead to deformation of the orthophoto patterns (as can be seen on the left side of the visible and infrared orthophotos in Figure~\ref{QR_Orthophotos_DM}). In case the DSM model is impacted, the depth map automatically undergoes the same deformation (as can be seen on the depth map in Figure~\ref{QR_Orthophotos_DM}). The second quality evaluation is the building of the depth map (Figure~\ref{QR_DM_Chessboard}). Despite the slight deformation in the left side of the parcel, the result of the depth map is consistent and well aligned with the visible orthophotos, and can be used in the segmentation process.

	In order to assess the added value of depth map information, two training sessions were performed on the SegNet~\cite{Badrinarayanan}, U-Net~\cite{Ronneberger}, DeepLabv3+~\cite{Chen2018a} and PSPNet~\cite{Zhao2017} networks. The first training session was conducted only on multispectral data, and the second one on multispectral data combined with depth map information. Figures~\ref{Difference_Between_Training_1} and~\ref{Difference_Between_Training_2} illustrate the qualitative test results of the comparison between the two trainings. The left side of Figure~\ref{Difference_Between_Training_1} shows an example of a parcel with a green ground. The center of the figure presents the segmentation result of the SegNet model trained only on multispectral data. As can be seen, in some areas of the parcel, it is difficult to dissociate vinerows. The right side of the figure depicts the segmentation result of the SegNet model trained on multispectral data combined with depth map information. This result is better than the previous one and it allows to easily separate vinerows. This is due to additional depth map information that allows a better learning of the scene environment and distinction between classes. Figure~\ref{Difference_Between_Training_1} illustrates other examples realised under the same conditions as above. On the first row, we observe an area composed of green ground. The segmentation results using the first and second models are displayed in the centre and on the right side, respectively. We can notice in this example a huge confusion between ground and healthy vine classes. This is mainly due to the fact that the ground color is similar to the healthy vine one. This problem has been solved by adding depth map information in the second model, the result of which is shown on the right side. The second row of Figure~\ref{Difference_Between_Training_2} presents an example of a partially diseased area. The first segmentation result reveals a detection of the disease class on the ground. The brown color (original ground color) merged with a slight green color (grass color) on the ground confused the first model and led it to misclassifying the ground. This confusion does not exist in the second segmentation result (right side). From these results, it can be concluded that the second model learned that the diseased vine class cannot be detected on "no-vine" when this one was trained on multispectral and depth map information. Based on these results, the following experiments were conducted using multispectral data and the depth map information.

	In order to validate the proposed architecture, a comparative study was conducted on the most well-known deep learning architectures, SegNet~\cite{Badrinarayanan}, U-Net~\cite{Ronneberger}, DeepLabv3+~\cite{Chen2018a} and PSPNet~\cite{Zhao2017}. All architectures were trained and tested on the following classes: shadow, ground, healthy and diseased with the same data (same training and test). Table~\ref{TabSegResult} lists the segmentation results of the different architectures. The quantitative evaluations are based on the F1-score and the global accuracy. As can be seen, the shadow and ground classes have obtained an average scores of 94\% and 95\% respectively with all architectures. The high scores are due to the easy detection of these classes. The healthy class has scored between 91\% and 92\% for VddNet, SegNet, U-Net and DeepLabv3+. However, PSPNet has obtained the worst result of 73.96\%, a score due to a strong confusion between the ground and healthy classes. PSPNet was unable to generate good segmentation model although the training dataset was rich. The diseased vine class is the most important class in this study. VddNet has obtained the best result for this class with a score of 92.59\%, followed by SegNet with a score of 88.85\%. The scores of the other architectures are 85.78\%, 81.63\% and 74.87\% for U-Net, PSPNet and DeepLabv3+ respectively. VddNet has achieved the best result because the features extraction was performed separately. Indeed, in~\cite{Kerkech2019} it has been proven that merging visible and infrared segmentations (with two separate trained models) provides a better detection than visible or infrared separately. The worst result of the diseased class was obtained with DeepLabv3+; this is due to a insensibility of the color variation. In fact, the diseased class can correspond to the yellow, brown or golden color and these colors are usually between the green color of healthy neighbour leaves. This situation has led classifiers to be insensitive to this variation. The best global segmentation accuracy was achieved by VddNet with an accuracy of 93.72\%. This score can be observed on the qualitative results of Figures~\ref{QR_Orthophoto_VddNet_1} and~\ref{QR_Orthophoto_VddNet_2}. Figure~\ref{QR_Orthophoto_VddNet_1} presents an orthophoto of a parcel (on the left side) partially contaminated with mildew. The right side shows the segmentation result by VddNet. It can be seen that it correct detects the diseased areas. Figure~\ref{QR_Orthophoto_VddNet_2} is an example of parcel without disease; here, VddNet also performs well performances in detecting true negatives.

	\begin{table*}[!h]
	\normalsize
	\renewcommand{\arraystretch}{1.25}
	\caption{\label{TabSegResult} Quantitative results with measurement of recall (Rec.), precision (Pre.), F1-Score~/~Dice (F1 / D.) and accuracy (Acc.) for the performances of VddNet, SegNet, U-Net, DeepLabv3+ and PSPNet networks, using multispectral and depth map data. Values are presented as a percentage.}	
	\setlength{\doublerulesep}{0pt}
	\begin{center}	
		
		\begin{tabular}{|c|c|c|c|c|c|c|c|c|c|c|c|c|c|} 
			\hline
			Class name	& \multicolumn{3}{c|}{Shadow} & \multicolumn{3}{c|}{Ground} & \multicolumn{3}{c|}{Healthy} & \multicolumn{3}{c|}{Diseased}& Total\\
			\hline
			Measure & Rec. & Pre. & F1/D. & Rec. & Pre. & F1/D. & Rec. & Pre. & F1/D. & Rec. & Pre. & F1/D. & Acc.\\
			\hline
			VddNet  & $ 94.88 $ & $ 94.89 $ & $ 94.88 $ & $ 94.84 $ & $ 95.11 $ & $ 94.97 $ & $ 87.96 $ & $ 94.84 $ & $ 91.27 $ & $ 90.13 $ & $ 95.19 $ & $ \textbf{92.59} $ & $ \textbf{93.72} $ \\
			\hline
			SegNet  & $ 94.97 $ & $ 94.60 $ & $ 94.79 $ & $ 95.16 $ & $ 94.99 $ & $ \textbf{95.07} $ & $ 90.14 $ & $ 94.81 $ & $ \textbf{92.42} $ & $ 83.45 $ & $ 95.00 $ & $ 88.85 $ & $ 92.75 $ \\
			\hline
			U-Net   &$ 95.09 $ & $ 94.70 $ & $ \textbf{94.90} $ & $ 94.99 $ & $ 95.07 $ & $ 95.03 $ & $ 89.09 $ & $ 94.74 $ & $ 91.83 $ & $ 78.27 $ & $ 94.90 $ & $ 85.78 $ & $ 90.69 $ \\			
			\hline
			DeepLabV3+  & $ 94.90 $ & $ 94.68 $ & $ 94.79 $ & $ 95.21 $ & $ 94.90 $ & $ 95.06 $ & $ 88.78 $ & $ 95.16 $ & $ 91.86 $ & $ 61.78 $ & $ 94.98 $ & $ 74.87 $ & $ 88.58 $ \\
			\hline			
			PSPNet  & $ 95.07 $ & $ 94.25 $ & $ 94.66 $ & $ 94.94 $ & $ 87.29 $ & $ 90.95 $ & $ 60.54 $ & $ 95.04 $ & $ 73.96 $ & $ 71.70 $ & $ 94.75 $ & $ 81.63 $ & $ 84.63 $ \\
			\hline	
		\end{tabular}
	\end{center}
\end{table*}

	\section{Conclusion}
	\label{Conclusion}
	The main goal of this study is to propose a new method that improve vine disease detection in UAV images. A new deep learning architecture for vine disease detection (VddNet), and automatic multispectral orthophotos registration have been proposed. UAV images in the visible and near-infrared spectra are the input data of the detection system for generating a disease map. UAV input images were aligned using an optimized multispectral registration algorithm. Aligned images are then used in the process of building registered orthophotos. During this process, a digital surface model (DSM) is generated to built a depth map. At the end, VddNet generates the disease map from visible, near-infrared and depth map data. The proposed method have brought many benefits to the whole process. The automatic multispectral orthophotos registration provides a high precision and fast processing compared to conventional procedures. A 3D processing enables the building of the depth map, which is a relevant data for VddNet training and segmentation process. Depth map data reduces misclassification and confusion between close color classes. VddNet improves disease detection and global segmentation compared to the state-of-the-art architectures. Moreover, orthophotos are georeferenced with GNSS coordinates, making it easier to locate diseased vines for traitment. For future work, it would be interesting to acquire new multispectral channels to enhance disease detection and improve the VddNet architecture.
	
	\section*{Acknowledgment}
	This work is part of the VINODRONE project supported by the Region Centre-Val de Loire (France). We gratefully acknowledge Region Centre-Val de Loire for its support.\\
	
	\bibliography{MyBibFile}
	
\end{document}